\documentstyle[aps,twocolumn,prb,epsf]{revtex}

\begin{document}
\draft

\twocolumn[\hsize\textwidth\columnwidth\hsize\csname@twocolumnfalse\endcsname
\title{ $T=0$ Phase Diagram of the Double-Exchange Model}  
\author{A.\ Chattopadhyay$^1$, A.\ J.\ Millis$^2$ and  S.\ Das Sarma$^1$} 
\address{ 
         $^1$ Department of Physics and MRSEC, University of Maryland\\
              College Park, MD 20742\\
         $^2$ Center for Materials Theory\\
              Department of Physics and Astronomy, Rutgers University\\       
              Piscataway, NJ 08854\\ }
\date{\today}
\maketitle

\widetext
\begin{abstract}
\noindent

We present the $T=0$ phase diagram of the double-exchange model 
(ferromagnetic Kondo lattice model) for all values of the carrier
concentration $n$ and Hund's couplng $J$, within dynamical mean field
theory. We find that depending on the values of $n$ and $J$, the ground
state is either a ferromagnet, a commensurate antiferromagnet or some other
incommensurate phase with intermediate wave vectors. The antiferromagnetic 
phase is separated by first order phase boundaries and wide regimes of phase 
separation. The transition from the ferromagnetic phase to an incommensurate
phase is second order. 

\end{abstract}

\vspace*{1cm}]

\narrowtext

\section{INTRODUCTION}

In this Brief Report we present the $T=0$ magnetic phase diagram of
the double exchange model\cite{Degennes60} believed relevant to the physics
of 'colossal' magnetoresistance manganites \cite{tokurabook}. Our results
may also be of interest in the context of magnetic semiconductors 
\cite{DasSarma,MacD}. We obtain the different magnetic phases, the boundaries
between them, and the regimes of phase separation using the dynamical mean
field approximation (DMFT) \cite{Amit00,Furukawa,Millis96b}. We consider
only the double-exchange interaction, and study only magnetic ordering, not
charge or orbital ordering. This paper is a companion to a previous paper
\cite{Amit00} which reported results mainly in regimes believed directly
relevant to experiment on manganites. Here we present the {\it complete} $T=0$
phase diagram of the model, for the convenience
of other workers, and to provide a basis for quantitative comparison of the
dynamical mean field method to other methods, i.e the phase diagram given 
in this paper is not just limited to CMR physics in manganites, but is of
much more general validity covering other systems like Mn doped GaAs
\cite{MacD}.

The double-exchange-only (DE) model involves electrons moving in a band
structure defined by a hopping matrix $t_{ij}^{ab}$ and a chemical potential 
$\mu $, and connected by a Hund's coupling ($J>0$) to core spins ${\bf
{S}}$. We denote the operator creating an electron in orbital $a$ with spin $%
\alpha $ on site $i$ by $d_{ia\alpha }^{\dagger }$ and define the
double-exchange only Hamiltonian $H_{DE}$ by\cite{Kubo72}  \begin{eqnarray}
H_{DE}= & - &\sum_{<ij>\alpha }t_{ij}^{ab}d_{ia\alpha }^{\dagger
}d_{jb\alpha }-\mu \sum_{i\alpha }d_{i\alpha }^{\dagger }d_{i\alpha } 
\nonumber \\
& - &J\sum_{i\alpha \beta }{\vec{S}}_{c,i}\cdot d_{i\alpha }^{\dagger }{\vec{%
\sigma}}_{\alpha \beta }d_{i\beta } \,.  
\label{Hdex}
\end{eqnarray}
where $<ij>$ implies sum over nearest neighbors. 

Note that there are only two 
independent dimensionless parameters defining $H_{DE}$, since the hopping
amplitude can be used to define the energy unit.
Our interest in this paper is the $T=0$ magnetic phase diagram of $H_{DE}$ as a
function of density $n$ and coupling $J$. 
To compute this we use the dynamical mean field approximation which becomes
exact in the limit of spatial dimensionality $d\rightarrow \infty $ and is
believed to be  reliable in $d=3 $ [Ref.\onlinecite{georges96}]. The complete
$T=0$ phase diagram of this important model has not to our knowledge been
presented in any limit. Other workers\cite{Furukawa,Millis96} have considered
two phases: a commensurate antiferromagnet (AF) and a homogeneous ferromagnet
(FM). In a previous paper\cite{Amit00} we demonstrated the existence of a free
energy minimum with spin order which is neither AF nor FM; this possible phase
has magnetic correlations incommensurate with the underlying lattice and was
denoted  IC. However, in our previous work the precise spin correlations were
not established, nor was the issue of  phase separation addressed.  The
importance of phase separation in the CMR context has been stressed by other
workers\cite {DagottoScience,Yunoki98}. Here, we resolve these issues. As
results obtained by other methods become more available, comparison to our
results will allow the accuracy of DMFT in $d=3$ to be assessed.

For our specific computations we take the core spins to be classical. This
is reasonable because we consider $T=0$ energetics in high spatial
dimension, and in cases of physical relevance the core spin is not small ($%
S=3/2$ for the manganites). We proceed by calculating the energies of the
different phases. We identify the ground state as the phase of lowest energy
and use an explicit Maxwell construction to determine regions of phase
separation.

We now discuss the possible phases which may occur. In the $d \rightarrow
\infty$ limit, Muller-Hartmann\cite{MH89} pointed out that the 
susceptibility depends on the one-dimensional parameter
$X(k)=\frac{1}{d}\sum_{i}cos(k_{i})$; similar considerations imply 
the ordering pattern and the energy are also functions of $X$. 
Because $-1 < X < 1$, we write $X=\cos(\theta)$; $\theta = 0$ corresponds 
to ferromagnetic (FM) phase and $\theta = \pi$ to antiferromagnetic 
(AFM) phase considered previously\cite{Amit00}. $0 < \theta < \pi$ 
coresponds to a phase with incommensurate (IC) spin correlations. The 
specific IC phase treated in our previous work\cite{Amit00} corresponds 
to $\theta= \pi/2$. 

The DMFT solution of Eq.~\ref{Hdex} may be written in terms of a 
spin-dependent mean field function ${\bf a}(\omega)$, which obeys an 
equation that depends on the density of states of the underlying lattice. 
We study the semicircular density of states 
$N(\epsilon)=\sqrt{4t^{2}-\epsilon ^{2}}/(2\pi t^{2})$, corresponding to the 
$d \rightarrow \infty$ limit of the Bethe lattice with nearest 
neighbor hopping, so that 
${\bf a}_i (\omega) = \omega + \mu - \sum_\delta t_{\delta} 
G^{loc}_{i+\delta,i+\delta} t_{\delta}$ with $t_{\delta}$ the hopping. To solve
the equations, we adopt a space dependent spin basis, quantized parallel
to the core spin on site $i$. The conventions just described then imply
that $\bf a$ has two independent components $a_{\uparrow}, a_{\downarrow}$ 
given by 
\begin{eqnarray}
a_{\uparrow}(\omega) &=&\omega +\mu - t^2\frac{\cos^{2}(\theta /2)}
{a_{\uparrow}(\omega)+J}-t^2\frac{\sin^{2}(\theta
/2)}{a_{\downarrow}(\omega)-J}   
\nonumber \\
a_{\downarrow}(\omega) &=&\omega +\mu -t^2\frac{\cos^{2}(\theta /2)}
{a_{\downarrow}(\omega)-J}-t^2\frac{\sin^{2}(\theta /2)}{a_{\uparrow}(\omega
)+J} 
\label{mfeq}
\end{eqnarray}

These equations contain as special cases the equations previously 
studied for the FM, AFM and $\theta = \pi/2$ phases. They may be obtained
by minimizing with respect to $a_{\uparrow}, a_{\downarrow}$, the energy
\begin{eqnarray}
E = T \sum_n \frac{[a(\omega_n)-(\omega_n + \mu)]^2}{t^2} + \frac{
b(\omega_n)^2}{t^2 \cos \theta} 
\nonumber \\
+ \ln [a(\omega_n)^2 - (b(\omega_n)+J)^2]
\end{eqnarray}
where $a = (a_{\uparrow}+a_{\downarrow})/2$ and $b =
(a_{\uparrow}-a_{\downarrow})/2$. This equation may be transformed into 
the familiar expression\cite{AGD}
\begin{eqnarray}
E &=&\int \,d\epsilon _{k}N(\epsilon _{k})\int_{-\infty }^{\infty
}\,\frac{d\omega}{\pi} f(\omega )\omega Tr[A(\epsilon _{k},\omega )]  \nonumber
\\ &=&\int_{-\infty }^{\infty }\,\frac{d\omega}{\pi} f(\omega )\omega Tr[%
\mathop{\rm Im}%
G_{loc}(\omega )]  
\label{energy}
\end{eqnarray}
through an integration by parts and use of the mean field equations
(recalling that $E$ is extremized by $a_{\uparrow},a_{\downarrow}$).
Here $N(\epsilon _{k})$ is the ''band'' density of states determined from
the hopping $t_{ij}^{ab}$, $A(\epsilon _{k},\omega )=-\frac{1}{\pi }
{\rm Im} G(\epsilon _{k},\omega )$ and the electron Green function ${\bf
G}=(\omega +\mu -\epsilon _{k}-{\bf \Sigma} (\omega ))^{-1}$. 
${\bf \Sigma } = {\bf a} - {\bf G}_{loc}^{-1}$ is the self
energy, which is local (k-independent) but dependent on spin in the dynamical
mean field approximation. Replacing the form of $G_{loc}$ in Eq.~\ref{energy}, 
we get 
\begin{equation}
E = \int \, \frac{d \omega}{\pi} f(\omega) \omega {\rm Im}\left
(\frac{2 a(\omega)}{a(\omega)^2- [b(\omega)+J]^2}\right)
\end{equation}

\begin{figure}[htbp]
\epsfxsize=3.5in
\epsfysize=3.75in
\epsffile{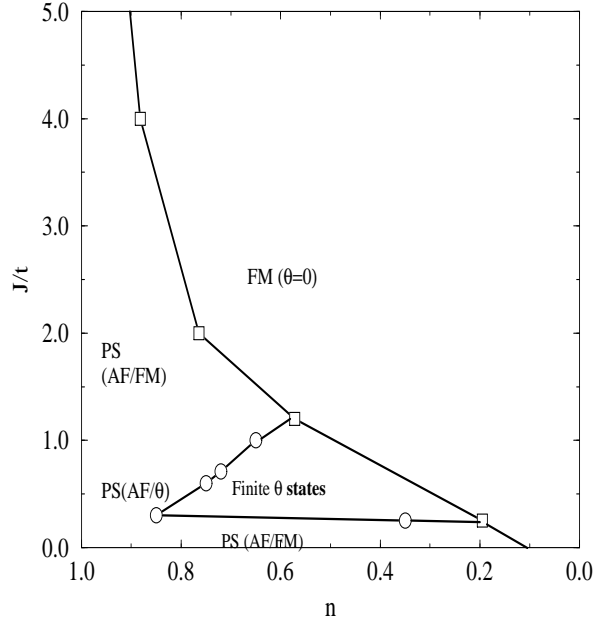}
\vspace{0.1in}
\caption{{\ The phase diagram of the double exchange model, as deduced from
the analytic calculations of the free energy at all $J$. }}
\label{phase}
\end{figure}

We have solved the mean field equations (Eq.~\ref{mfeq}) and computed the
energies for a wide range of $\theta,n$. The resulting phase diagram is shown
in Fig.~\ref{phase}. As one can see from Fig.~\ref{phase}, a homogeneous
ferromagnetic phase is supported over a wide range of fillings and couplings.
There is a small island of IC phase characterized by a $\theta$ which varies
with position inside the IC phase. The AFM phase exists only at $n = 1$. 
The FM-IC phase boundary is apprently second order; all of the other phase
boundaries are (within our numerical accuracy) first-order and are accompanied
by regions of phase separation. At sufficiently small $J$, phase separation
preempts the IC phase entirely even though this phase minimizes the energy for
$0.47\leq n\leq 0.7$.

\begin{figure}[htbp]
\epsfxsize=3.3in
\epsfysize=3.0in
\epsffile{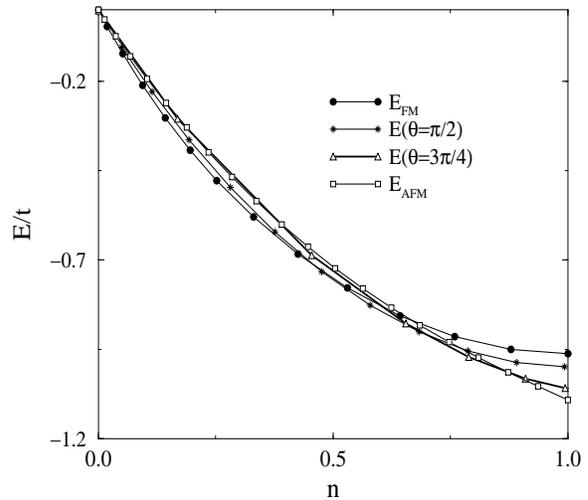}
\vspace{0.1in}
\caption{{ Energy per orbital for $J/t=0.6$. There is a continuity of
phases from $\theta=0$ to $\theta = 3 \pi/4$, followed by phase separation
between $\theta=3 \pi/4$ and $\theta = \pi$ }}
\label{J0.6}
\end{figure}

\begin{figure}[htbp]
\epsfxsize=3.3in
\epsfysize=3.0in
\epsffile{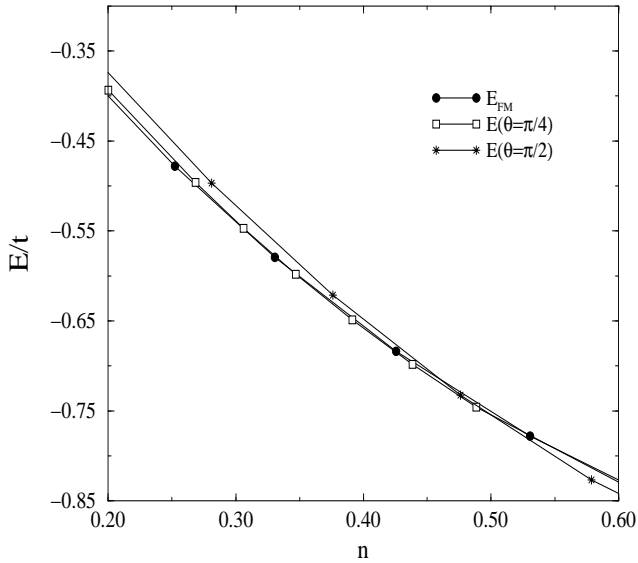}
\vspace{0.1in}
\caption{{The region between $0.2\le n \le 0.6$ has been amplified to 
illustrate that the transitions between $\theta=0$ to $\theta = \pi/2$ 
are second order. The second order transitions extend to the $\theta = 
3\pi/4$ state}} 
\label{J0.6_2}
\end{figure}

Fig.~\ref{J0.6} shows energy as a function of $n$ for the FM and AF 
states, as well as $\theta= \pi/2$ and $3 \pi/4$. We see immediately from 
these curves that at $n=1$ the AF phase is the lowest; at intermediate
$n$ there is a sequence of IC phases characterized by a varying angle 
and at small $n$ the FM phase is stable. Between the $n=1$ AF phase and 
the intermediate $n$ IC phase there is a region of phase separation, and 
the $\theta$ value at the PS-IC boundary is a non-universal $J$-dependent 
number. Fig.~\ref{J0.6_2} shows the behavior in the vicinity of the FM/IC 
phase boundary. We see that the transition is apprently second-order, with
$\theta$ evolving smoothly to 0 at the FM/IC phase boundary.  

\begin{figure}[htbp]
\epsfxsize=3.3in
\epsfysize=3.0in
\epsffile{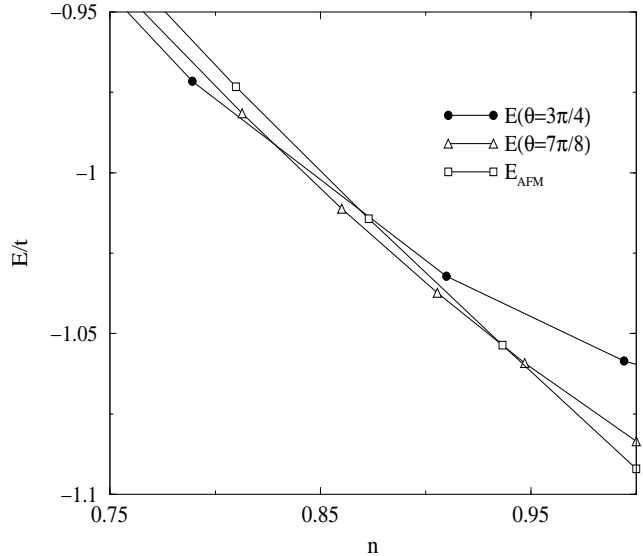}
\vspace{0.1in}
\caption{{ The transition from the $\theta = 3 \pi/4$ state to the AFM state
is first order, and the system phase separates.}} 
\label{J0.6_3}
\end{figure}

Fig.~\ref{J0.6_3} shows the behavior for $J=0.6 t$ and $n$ close to $1$. A 
Maxwell construction shows that the $\theta=7\pi/8$ curve lies higher 
than the line connecting the AF $n=1$ point and the $\theta=3\pi/4, 
n \approx 0.78$ point. The energy differences involved are seen to be tiny, 
implying large uncertainties in the location of the phase boundary and 
rather complicated behavior in physical systems. 

A generally noteworthy feature of Fig.~\ref{J0.6_2} is
the extremely small energy differences between different phases: typically
the energy difference per site $\Delta E$ between the lowest and next lowest
phase is only about $0.02t$. For manganites, a reasonable value of $t$ is
$0.65$eV \cite{Amit00} leading to $\Delta E/site \sim 100 K$. This suggests
an unusual sensitivity to disorder, as already noted by Moreo et al.\cite
{Moreo99}.

We now use analytic arguments to analyze more carefully the behavior in the
small $J$ limit, where the energy may be written as 
\begin{equation}
E = -\frac{1}{2} \chi(\theta) J^2
\label{smallJ}
\end{equation}
with 
\begin{equation}
\chi(\theta) = -2 \int \, \frac{d \omega}{\pi} f(\omega) {\rm Im}\left(\frac{1}
{a_0(\omega)^2- t^2 \cos \theta}\right)
\label{chitheta}
\end{equation}
and $a_0 = \frac{1}{2}[\omega + \mu - \sqrt{(\omega + \mu)^2 - 4 t^2}]$

\begin{figure}[htbp]
\epsfxsize=3.3in
\epsfysize=3.0in
\epsffile{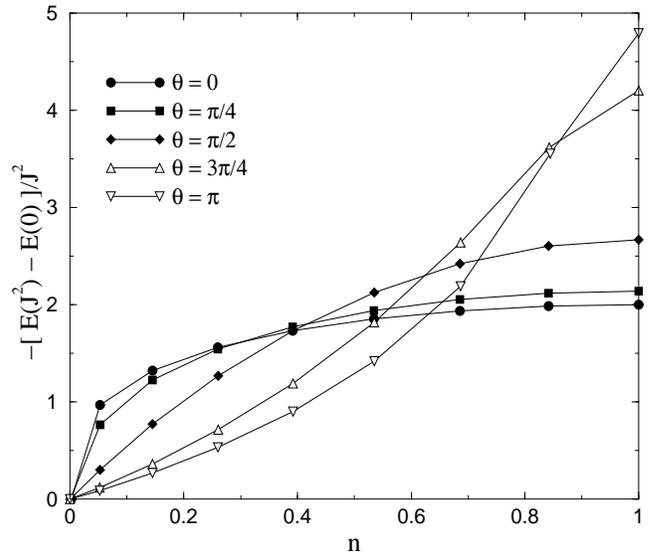}
\vspace{0.1in}
\caption{{\ Small $J$ energies of the states with different $\theta$ computed
from Eqs.~\ref{chitheta} and \ref{chiaf} at $J=0.05t$. The wide regime of
phase separation between $\theta=0, n= 0.15$ and $\theta=\pi, n=1$ phases
is evident.
}} \label{chi}
\end{figure}

Note that the behavior as $\theta \rightarrow \pi$ and $n \rightarrow 1$
is complicated because the energy of the $n=1, \theta=\pi$ AFM phase 
$\sim J^2 \ln(1/J)$. From the explicit energy expression we find 
(neglecting $O(J^4)$)
\begin{eqnarray}
\frac{\partial E_{AF}}{\partial J^{2}} &=&\frac{J}{4}\left[ E\left(
sin^{-1}(\mu /2); -4/J^{2}\right) -E\left( \pi /2; -4/J^{2}\right) \right] - 
\nonumber \\
&&\frac{4+J^{2}}{4J}\left[ F\left( sin^{-1}(\mu /2); -4/J^{2}\right) -F\left(
\pi /2; -4/J^{2}\right) \right]   
\label{chiaf}
\end{eqnarray}
where $F(\phi ;m),E(\phi ;m)$ are elliptic integrals of the First and Second
Kind. The corresponding energies are plotted in Fig.~\ref{chi} for $J=0.05$.
Note that for $n$ very near 1 ($\mu <J$), $\chi _{AF}$ is linear in $1-n$.
Thus at $n\neq 1$ the AF phase is always unstable to phase separation. The
Maxwell construction also shows that in the small $J$ limit the IC phase is
preempted entirely, and the ferromagnetic phase is only stable for $n\leq
0.15$.

We now briefly discuss the physical  meaning of the IC phase. From 
Eq.~\ref{smallJ}, the lowest energy phase is the one with the largest 
$\chi(\theta)$. For typical band structures even in $d=3$ $\chi $ takes one
value at $q=0$ and another value at $q=Q=(\pi ,\pi ,\pi );$ the variation
between these limits is smooth for momenta along the zone diagonal, but over
much of the Brillouin zone $\chi $ is quite weakly momentum dependent.  In the
$d\rightarrow \infty $ limit this weak momentum dependence disappears: $\chi
(\vec{q})$ at any $\vec{q} \neq \lambda (\pi ,\pi ,...)$ takes the
$q$-independent value $\chi _{loc}=\sum_{q}\chi (\vec{q})$ and along the
diagonal. For $d$ large but finite,  $\chi $ varies rapidly with $q$
crossing over from its $\vec{q}=\lambda (\pi ,\pi ,...)$ values to a
very weakly $q$-dependent value in a range $\delta q\approx O(1/\sqrt{d})$.
Thus the hallmark of the phases which are characterized by a small 
$\cos \theta $ is a large range of spin arrangements very nearby in
energy. We suspect that this regime would appear experimentally as a spin
glass in the presence of disorder 

In conclusion, we have determined the $T=0$ phase diagram of the
double-exchange Hamiltonian including regions of phase separation, using
analytic arguments at small and large $J$ and the dynamical mean field method.
We find that the incommensurate phase noted in Ref.~\cite{Amit00} has a rather
small region of existence, being mostly overwhelmed by phase separation, and
have found the differences in energy between the different phases to be very
small.

We have learned of related work by L.\ Yin and J.\ Ho (APS March
Meeting 2000 Bulletin; L.\ Yin and J.\ Ho, private communication), who used
a different mean-field technique in $d=3$ to obtain a phase diagram
including spiral states, with regions of phase separation qualitatively
similar to what is found here. Our results are also similar to the 
phase diagram of Alonso et al.\cite{Guinea}.

{\it Acknowledgements} We thank J.\ K.\ Freericks for suggesting
that we examine the $\theta$ dependence of the incomensurate phase
and the University of Maryland NSF-MRSEC, NSF-DMR-0081075 (AJM) and
US-ONR (SDS) for support.

\end{document}